\documentclass[sigconf]{acmart}

\usepackage{amsmath,graphicx}
\usepackage{multirow}
\usepackage{epstopdf}
\usepackage{rotating}
\usepackage{mathrsfs}

\usepackage{amssymb}

\usepackage[caption=false]{subfig}
\usepackage{algorithm}
\usepackage{algorithmic}
\usepackage{xcolor}
\usepackage{makecell}
\usepackage{array}
\usepackage{hyperref}
\usepackage{balance}
\AtBeginDocument{%
  \providecommand\BibTeX{{%
    \normalfont B\kern-0.5em{\scshape i\kern-0.25em b}\kern-0.8em\TeX}}}

\copyrightyear{2020}
\acmYear{2020}
\setcopyright{acmcopyright}\acmConference[MM '20]{Proceedings of the 28th ACM International Conference on Multimedia}{October 12--16, 2020}{Seattle, WA, USA}
\acmBooktitle{Proceedings of the 28th ACM International Conference on Multimedia (MM '20), October 12--16, 2020, Seattle, WA, USA}
\acmPrice{15.00}
\acmDOI{10.1145/3394171.3413839}
\acmISBN{978-1-4503-7988-5/20/10}

\begin{document}

\fancyhead{}

\title{Unpaired Image Enhancement with Quality-Attention Generative Adversarial Network}


\author{Zhangkai Ni}
\orcid{0000-0003-3682-6288}
\affiliation{%
  \institution{City University of Hong Kong}
  \city{Hong Kong}
}
\email{eezkni@gmail.com}

\author{Wenhan Yang}
\affiliation{%
 \institution{City University of Hong Kong}
 \city{Hong Kong}
}
\email{yangwenhan@pku.edu.cn}

\author{Shiqi Wang}
\authornote{Corresponding authors: Sam Kwong and Shiqi Wang}
\affiliation{%
 \institution{City University of Hong Kong}
 \city{Hong Kong}
}
\email{shiqwang@cityu.edu.hk}

\author{Lin Ma}
\affiliation{%
 \institution{Meituan-Dianping Group}
 \city{China}
}
\email{forest.linma@gmail.com}

\author{Sam Kwong}
\authornotemark[1]
\affiliation{%
 \institution{City University of Hong Kong}
 \city{Hong Kong}
}
\email{cssamk@cityu.edu.hk}


\renewcommand{\shortauthors}{Ni, et al.}

\begin{abstract}
  In this work, we aim to learn an unpaired image enhancement model, which can enrich low-quality images with the characteristics of high-quality images provided by users. We propose a quality attention generative adversarial network (QAGAN) trained on~\textit{unpaired} data based on the bidirectional~\textit{Generative Adversarial Network} (GAN) embedded with a~\emph{quality attention module} (QAM). The key novelty of the proposed QAGAN lies in the injected QAM for the generator such that it learns domain-relevant quality attention~\textit{directly} from the two domains. More specifically, the proposed QAM allows the generator to effectively select semantic-related characteristics from the spatial-wise and adaptively incorporate style-related attributes from the channel-wise, respectively. Therefore, in our proposed QAGAN, not only discriminators but also the generator can directly access both domains which significantly facilitate the generator to learn the mapping function. Extensive experimental results show that, compared with the state-of-the-art methods based on unpaired learning, our proposed method achieves better performance in both objective and subjective evaluations.
\end{abstract}




\begin{teaserfigure}
	\centering
	\centerline{\includegraphics[width=1.0\linewidth]{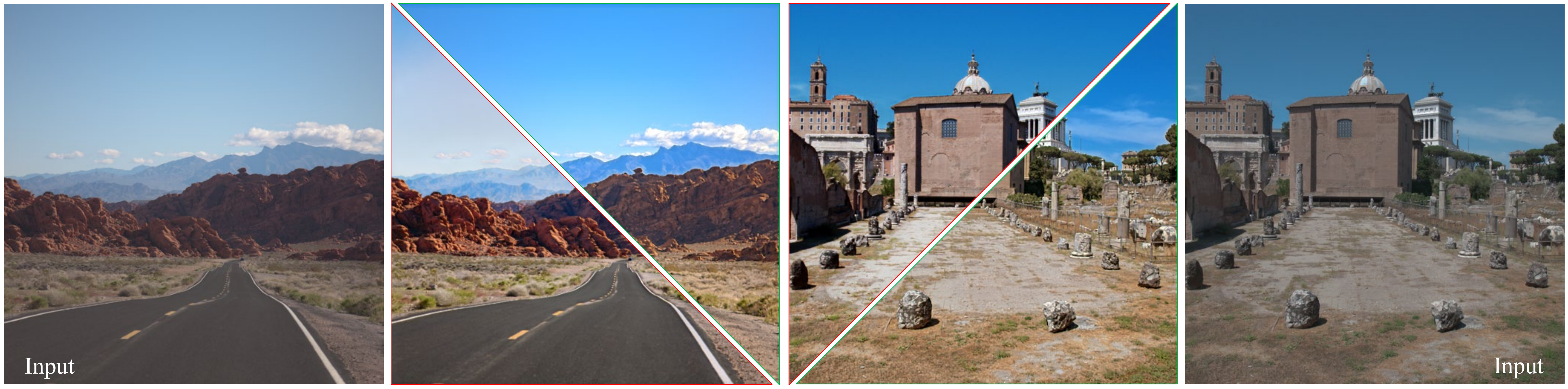}}
	\caption{Examples of our proposed QAGAN model trained on two different unpaired datasets. (\textit{i.e.}, results of models trained using the MIT-Adobe FiveK dataset (red) and the crawled Flickr dataset (green).)}
	\label{fig:framework}
\end{teaserfigure}

\maketitle

\section{Introduction}

Image enhancement aims to improve the quality of images via various post-processing techniques, such as contrast enhancement, color rendition, and sharpening, to name a few. Experienced photographers can freely generate their favorite visual pleasing images through the professional image-editing software (\textit{e.g.}, Adobe Photoshop and Lightroom), which are desired by the general public who lacks professional image-editing skills. This contradiction highlights the importance of the~\textit{user-oriented} automatic image enhancement method for the general public to produce high-quality images they want. Furthermore, automatic image enhancement is already a build-in technology for displays, cameras, scanners, and photography applications to provide users with better customized services.

Significant progress has been made in improving the performance of automatic image enhancement methods~\cite{bychkovsky2011learning, gharbi2017deep, ignatov2017dslr, yan2016automatic, ren2019low, wang2019underexposed}. However, most of these models require a large-scale dataset with~\textit{paired} low/high-quality images in order to learn the enhancement mappings in a~\textit{full-supervised} manner. In general, such paired training images can either be manually edited by experienced photographers to generate corresponding high-quality counterparts from low-quality images~\cite{bychkovsky2011learning}, or both kinds of images are captured simultaneously and well-registered to ensure the pixel correspondence~\cite{ignatov2017dslr}. Numerous efforts have been devoted to collecting the low/high-quality training image pairs~\cite{bychkovsky2011learning, ignatov2017dslr, wang2019underexposed} to take advantage of data-driven learning to train automatic image quality enhancement models. Unfortunately, such supervised models have the following two drawbacks. On the one hand, it is difficult to collect paired training images in dynamic scenes with various motions. On the other hand, even if such paired images are collected, they cannot reflect the user preferences and cannot play the role in guiding the training/learning of the enhancement model. The user preference is a kind of property that supervised learning methods hardly possess. Since the perception of visual quality is a cognitive task that is highly relevant to the human personality, educational background, and aesthetic standard, whereas the quality of carefully prepared paired training images are usually highly dependent on the creator (\textit{e.g.}, retoucher or photographer).

In this study, we make efforts in addressing the unpaired image enhancement task that only needs to be supervised by the unpaired image set (without any paired images) expressing personal user preferences. The nature of such unpaired learning makes it possible to easily create~\textit{personalized} training data for everyone to learn their own~\textit{individual-}\textit{oriented} image enhancement model. Our proposed method is based on a bidirectional generative adversarial network (GAN)~\cite{goodfellow2014generative} between the source domain (\textit{i.e.}, low-quality images domain) and the target domain (\textit{i.e.}, high-quality images domain). However, the generator of most existing GAN-based models can only learn the features of the target domain under the guidance of discriminator because only the discriminator can access both domains. This increases the burden of learning the ideal mapping between different domains and also leads to instability in GAN training. To overcome this difficulty, we propose a~\textit{quality attention module} (QAM) which allows the generator to effectively select semantic-relevant characteristics from the spatial-wise and adaptively incorporate style-related natures from the channel-wise, respectively. In other words, in our proposed QAGAN model, not only the discriminator but also the generator can directly access both two domains, which significantly facilitates the generator to learn a more efficient mapping function that encodes richer context. This is the significant difference between our proposed QAGAN with the well-known CycleGAN~\cite{zhu2017unpaired} and the related follow-up works. In particular, we employ the VGG-19 network~\cite{simonyan2014very} as the encoder of the generator, such that the entire framework of our algorithm only needs to learn two decoders and two corresponding discriminators with unpaired images. The main contributions of our work can be summarized as follows:
\begin{itemize}
\item We propose a bidirectional QAGAN framework embedded with QAM which performs~\textit{user-oriented} unpaired image enhancement by feed-forward learning.
\item We propose a~\textit{quality attention module} (QAM) that enables the generator to directly learn the domain-relevant quality adaptively rather than only under the guidance of the discriminator.
\item Extensive experiments demonstrate the superiority of the proposed method compared to the state-of-the-art in qualitative and quantitative evaluations.
\end{itemize}

\begin{figure*}[t]
	\centering \centerline{\includegraphics[width=1.0\linewidth]{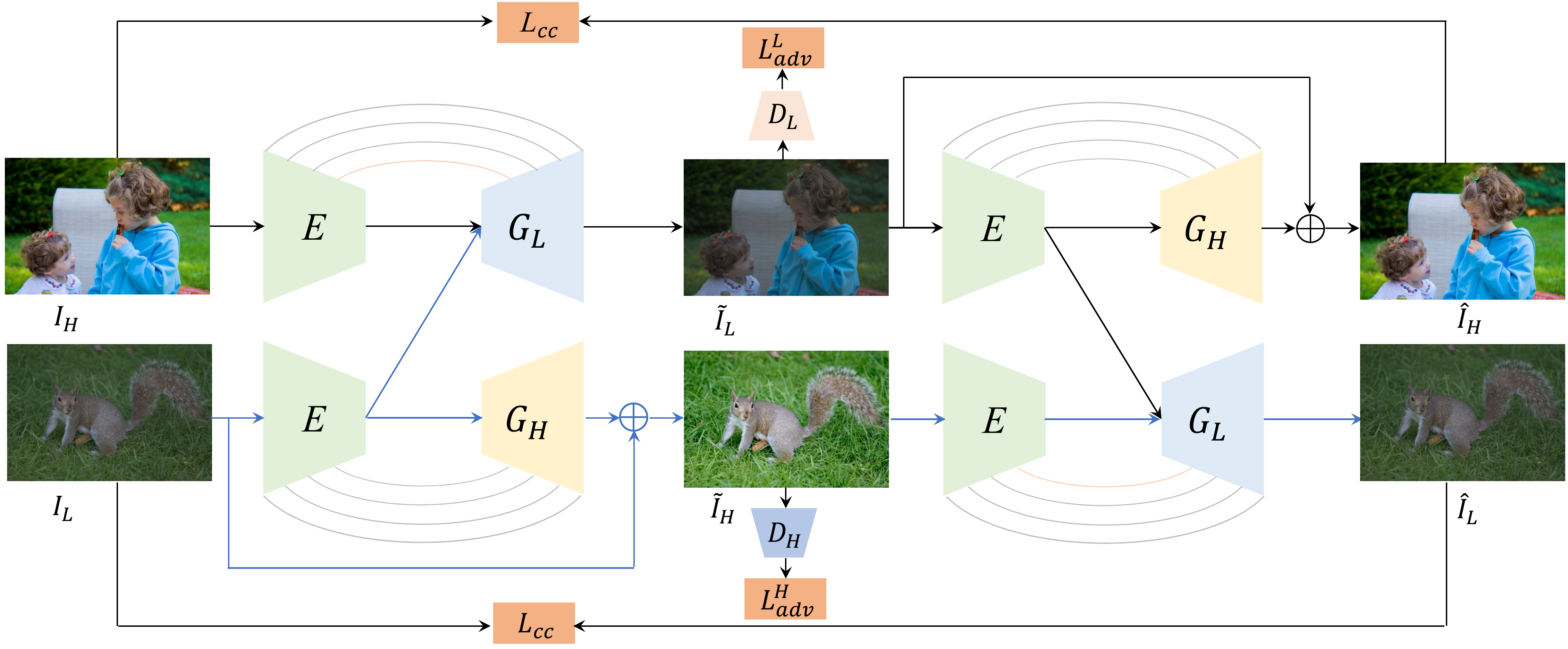}}
	\caption{An overview of the proposed Quality Attention Generative Adversarial Network (QAGAN) for image enhancement. $E$ is a pre-trained encoder (\textit{i.e.}, VGG-19~\cite{simonyan2014very} in our work) used to extract image features. $G_L$ and $G_H$ are low-quality and high-quality image generators, respectively. The discriminator $D_L$ and $D_H$ try to distinguish the generated low-quality image $\tilde{I}_L$ from real low-quality images, and distinguish the generated high-quality image $\tilde{I}_H$ from the real high-quality ones. The proposed framework uses the adversarial loss $L_{adv}$ and cycle-consistency loss $L_{cc}$ to train from the unpaired data in an end-to-end manner.}
	\label{fig:framework}
\end{figure*}

\section{Related Work}

\subsection{Automatic Image enhancement}

\textbf{\quad Conventional Image Enhancement:} Extensive efforts have been made over the past few decades to automatically improve the aesthetic quality of images. Conventional image enhancement algorithms can be roughly divided into the following three categories:~\textit{histogram equalization}~\cite{thomas2011histogram, coltuc2006exact},~\textit{unsharp masking}~\cite{ye2018blurriness}, and~\textit{Retinex-based methods}~\cite{li2018structure, guo2017lime}. Most of these methods either extract pre-defined sophisticated handcrafted features or adjust the image locally or globally based on prior knowledge. In addition, most of these adjustments are only used for individual types of improvement, such as contrast enhancement, edge sharpening, or color adjustment. For instance, Thomas~\textit{et al.}~\cite{thomas2011histogram} adjusted the luminance histogram to a specific distribution based on the statistical information of the entire image to stretch the contrast. Ye~\textit{et al.}~\cite{ye2018blurriness} adaptively performed pixel-wise enhancement under the guidance of blurriness to sharpen the image. Moreover, some works tried to manipulate image from visual attention or saliency~\cite{wong2011saliency, nguyen2013image, mateescu2015visual, mechrez2019saliency}. Such as, Nguyen~\textit{et al.}~\cite{nguyen2013image} proposed an image re-attentionizing framework to make the target region in the image attract the attention of humans. Mechrez~\textit{et al.}~\cite{mechrez2019saliency} designed an optimization framework to change visual saliency based on internal patch statistics.

\textbf{Data-driven Image Enhancement:} Due to the excellent modeling capabilities of deep convolutional neural network (CNN), the state-of-the-art image enhancement algorithms are almost all based on data-driven models~\cite{gharbi2017deep, ignatov2017dslr, yan2016automatic, wang2019underexposed, ren2019low, chen2018deep, jiang2019enlightengan, hu2018exposure, deng2018aesthetic}. Most of these methods, however, heavily depend on fully supervised learning that requires a large number of training image pairs. For example, Ignatov~\emph{et al.}~\cite{ignatov2017dslr} presented a model for improving the image quality of mobile devices to make it close to that of high-quality digital single-lens reflex photos. In order to enable the model to learn in a supervised manner, it takes a lot of time and efforts to create a paired training dataset DSLR Photo Enhancement Dataset (DPED). Gharbi~\emph{et al.}~\cite{gharbi2017deep} realized the real-time image enhancement by jointly considering the local and global information to dynamically enhance the image. Wang~\emph{et al.}~\cite{wang2019underexposed} proposed an image-to-illumination mapping model trained from a newly established under-exposed image dataset that consists of 3000 pairs of under-exposed images and corresponding retouched versions. However, the practicality of all these models is limited by the supervised models that rely heavily on the paired training data.

In recent years, the development of reinforcement learning (RL) and generative adversarial network (GAN)~\cite{goodfellow2014generative} has brought significant progress to data-driven image processing, especially the~\textit{image-to-image translation} tasks with~\textit{unpaired learning}. Hu~\textit{et al.}~\cite{hu2018exposure} proposed an RL-based framework that learns retouched images similar to those edited by professional photographers with sequentially apply differential filters. Kosugi~\textit{et al.}~\cite{kosugi2019unpaired} developed an RL-based model to handle image enhancement by directly using image editing software (\textit{e.g.}, Adobe Photoshop). Deng~\textit{et al.}~\cite{deng2018aesthetic} proposed a model based on adversarial learning, which is trained under the supervision of the aesthetic quality of binary tags. One of the most classical models dedicated to unpaired learning is the CycleGAN~\cite{zhu2017unpaired}, which uses cycle consistency and adversarial losses to learn the transform mappings between two domains. The model is concise and explicit, but it makes unpaired image processing possible and achieve relative good results. Beyond CycleGAN, Chen~\textit{et al.}~\cite{chen2018deep} made some improvements to stable the training of GAN to achieve better enhanced results. However, in these works, only the~\textit{discriminators} can access both domains while the~\textit{generators} can only learn the corresponding unidirectional mapping under the guidance of discriminators. On the contrary, the generator in our proposed model can also access both domains to directly learn the domain-relevant quality features which benefit the learning procedure significantly.

\subsection{Attention Mechanism}

Recently, attention mechanisms have been extensively studied for video classification~\cite{wang2018non}, image generation~\cite{zhang2018self}, and image restoration~\cite{liu2018non}. The spatial attention mechanism for computer vision is proposed to make up for the defect that CNN is ineffective in capturing the long-range dependencies due to the locality of the convolution operations. Wang~\textit{et al.}~\cite{wang2018non} formulated the attention mechanism as the~\textit{non-local} operation to explore long-range dependencies. Zhang~\textit{et al.}~\cite{zhang2018self} demonstrated that the quality of the generated image can be significantly improved by modeling the self-attention within an image. Liu~\textit{et al.}~\cite{liu2018non} proposed a non-local recurrent neural network (RNN) to use self-similarity prior in images for image denoising. All these attention models share the same idea, that is, correlation among non-local regions within an image/video is an effective prior for image processing. Our proposed~\textit{quality attention module} (QAM) is very close to this idea but we take advantage of the similarity between images from different domains.

\section{Method}

An overview of the proposed unpaired learning model for image enhancement is shown in Fig.~\ref{fig:framework}. Our proposed model is based on~\textit{bidirectional} GANs, which contains two generators: $G_H$ and $G_L$ are used to generate high-quality and low-quality images, respectively,
and are paired with two discriminators $D_H$ and $D_L$, respectively. The encoder $E$ and generator $G_H$ form a U-Net~\cite{ronneberger2015u} structure dedicated to translating the low-quality image into a high-quality version. The generator $G_L$ embeds the proposed~\textit{quality attention module} (QAM) to directly learn the domain-relevant quality features from the features of low-quality images. The discriminators $D_L$ and $D_H$ aim for distinguishing the generated data from the real low-quality images and high-quality images, respectively. The key novelty of the proposed QAGAN model lies in that we exploit the quality attention module of generator $G_L$ to directly learn domain-relevant quality attention from the low-quality image domain. The proposed attention mechanism significantly benefits the generator $G_L$ to achieve excellent performance in image mapping task from high-quality to low-quality, which therefore naturally leads to better performance in low-quality to high-quality image translation hereafter.

\subsection{Quality Attention Module}
\label{ssec:module}

Inspired by recent works on non-local self-similarity~\cite{wang2018non}, we make the first attempt to introduce the non-local similarity of images from different domains into the generator in bidirectional GANs to achieve unpaired image enhancement. The proposed~\textit{quality attention module} (QAM) allows the generator $G_L$ to effectively select semantic-relevant characteristics from the spatial-wise, and adaptively incorporate style-related attributes from the channel-wise, respectively. Therefore, the generator $G_L$ can significantly be benefited from the QAM to translate the high-quality image to the low-quality version with learned domain-relevant quality features.

\begin{figure}[t]
  \centering
  \includegraphics[width=0.98\linewidth]{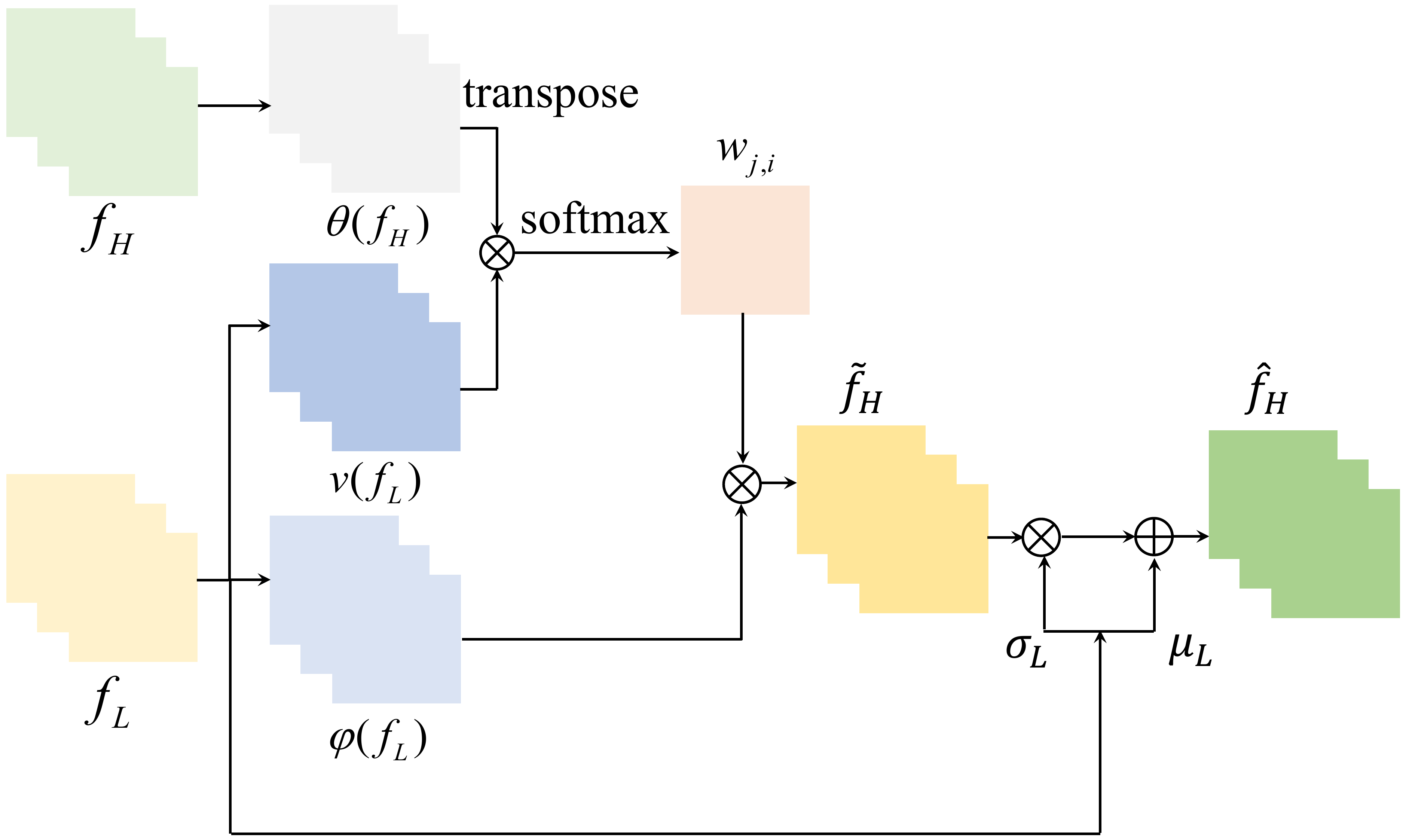}
  \caption{The structure of~\textit{quality attention module} (QAM). The QAM enables $f_H$ to adaptively select content-relevant features from the spatial non-local inter-domain similarity and then incorporate the style attributes from the channel-wise statistics.}
\label{fig:qam}
\end{figure}

The detailed structure of the proposed QAM is shown in Fig.~\ref{fig:qam}. In short, for given encoder features $f_L\in \mathbb{R}^{N\times C\times H\times W}$  of low-quality image $I_L$ and encoder features $f_H\in \mathbb{R}^{N\times C\times H\times W}$ of high-quality image $I_H$. The proposed QAM first uses non-local inter-domain spatial similarity to generate spatially weighted features $\tilde{f}_H\in \mathbb{R}^{N\times C\times H\times W}$ and then derives the channel-wise recalibration features $\hat{f}_H\in \mathbb{R}^{N\times C\times H\times W}$ according to the style of $f_L$, where $N$ is the value of batch size, $C$ is the number of channels, $H$ and $W$ are high and width of the features, respectively. More specifically, we first generate two new features $\nu(f_L)$ and $\varphi(f_L)$ from $f_L$ to reduce computational complexity, where $\{\nu(f_L), \varphi(f_L)\}\in \mathbb{R}^{N\times\frac{C}{8}\times H\times W}$. In a similar way, the $\theta(f_H)\in \mathbb{R}^{N\times\frac{C}{8}\times H\times W}$ is generated from $f_H$. After that, we reshape $\{\theta(f_H), \nu(f_L)\}\in \mathbb{R}^{N\times\frac{C}{8}\times H \times W}$ to $\{\theta(f_H), \nu(f_L)\}\in \mathbb{R}^{N\times\frac{C}{8}\times S}$, and then multiply $\nu(f_L)$ and the transpose of $\theta(f_H)$, where $S=H*W$. The inter-domain non-local spatial similarity map $w_{j,i}$ can be derived by:
\begin{equation}
\label{equ:sa}
w_{j,i}=\frac{e^{\theta(f_H)_i^T \cdot \nu(f_L)_j}}{\sum\nolimits_{i=1}^Se^{\theta(f_H)_i^T \cdot \nu(f_L)_j}},
\end{equation}
where $w_{j,i}$ represents the dependence of the $j$-th position of $f_L$ on the $i$-th position of $f_H$. The output of the inter-domain non-local spatial similarity can be formulated as the weighted sum of the features over all positions as follows:
\begin{equation}
\label{equ:saoutput}
e_i = \sum\nolimits_{i=1}^N(w_{j,i} \cdot \varphi(f_L)).
\end{equation}
The refined features $\tilde{f_H}$ can be obtained by simply reshaping the $e_i$. This inter-domain non-local spatial similarity allows the generator $G_L$ to effectively learn the spatial content-relevant characteristics from $I_L$ directly.

Inspired by~\cite{huang2017arbitrary}, we adjust the channel-wise statistics (\textit{i.e.}, mean and variance) of $I_H$ to match that of $I_L$ to inject the style of the low-quality image $I_L$ to the generated low-quality version $\tilde{I}_L$. We adopt this strategy because its efficiency and effectiveness have been witnessed in many style transfer works. Specifically, the style features mean $\mu_L$ and variance $\sigma_L$ can be computed as:
\begin{equation}
\small
\label{equ:mean}
\mu_L = \frac{1}{HW}\sum\limits_{h=1}^H\sum\limits_{w=1}^W f_{L},
\end{equation}

\begin{equation}
\label{equ:var}
\small
\sigma_L = \left(\frac{1}{HW}\sum\limits_{h=1}^H\sum\limits_{w=1}^W \left(f_{L}-\mu_L\right)^2\right)^{\frac{1}{2}},
\end{equation}
where, the $\mu_L$ and $\sigma_L$ are regarded as the style information of the low-quality image $I_L$. Therefore, we can translate it to the generated low-quality image $\tilde{I}_L$ by simply scaling the normalized $\tilde{f}_H$ with $\sigma_L$ and adding the bias $\mu_L$, which can be expressed as:

\begin{equation}
\label{equ:var}
\small
\hat{f}_H = \sigma_L \left(\frac{\tilde{f}_H-\mu_H}{\sigma_H}\right) + \mu_L,
\end{equation}
where $\mu_H$ and $\sigma_H$ are the mean and variance of $\tilde{f}_H$, respectively, and the calculation is similar to that of $f_L$.

In summary, the proposed QAM allows the generator $G_L$ to learn domain-relevant quality features from $I_L$ in terms of the spatial-wise based on non-local inter-domain similarity and channel-wise according to mean and variance.

\subsection{Network Architecture}
\label{ssec:network}

As shown in Fig.~\ref{fig:framework}, our framework employs the pre-trained VGG~\cite{simonyan2014very} as the encoder $E$ to transfer the $I_L$ and $I_H$ into the feature space. The generator $G_L$ is a symmetric decoder incorporated with the proposed QAM to convert the low-quality images into high-quality versions. The generator $G_H$ and $E$ are combined to form a U-Net to map the low-quality images into the high-quality counterparts. The associated discriminators $D_L$ and $D_H$ are responsible for distinguishing the generated low-quality image $\tilde{I}_L$ from the real low-quality images and judging the generated high-quality image $\tilde{I}_H$ from the real high-quality ones.

\begin{figure}[t]
  \centering
  \includegraphics[width=0.98\linewidth]{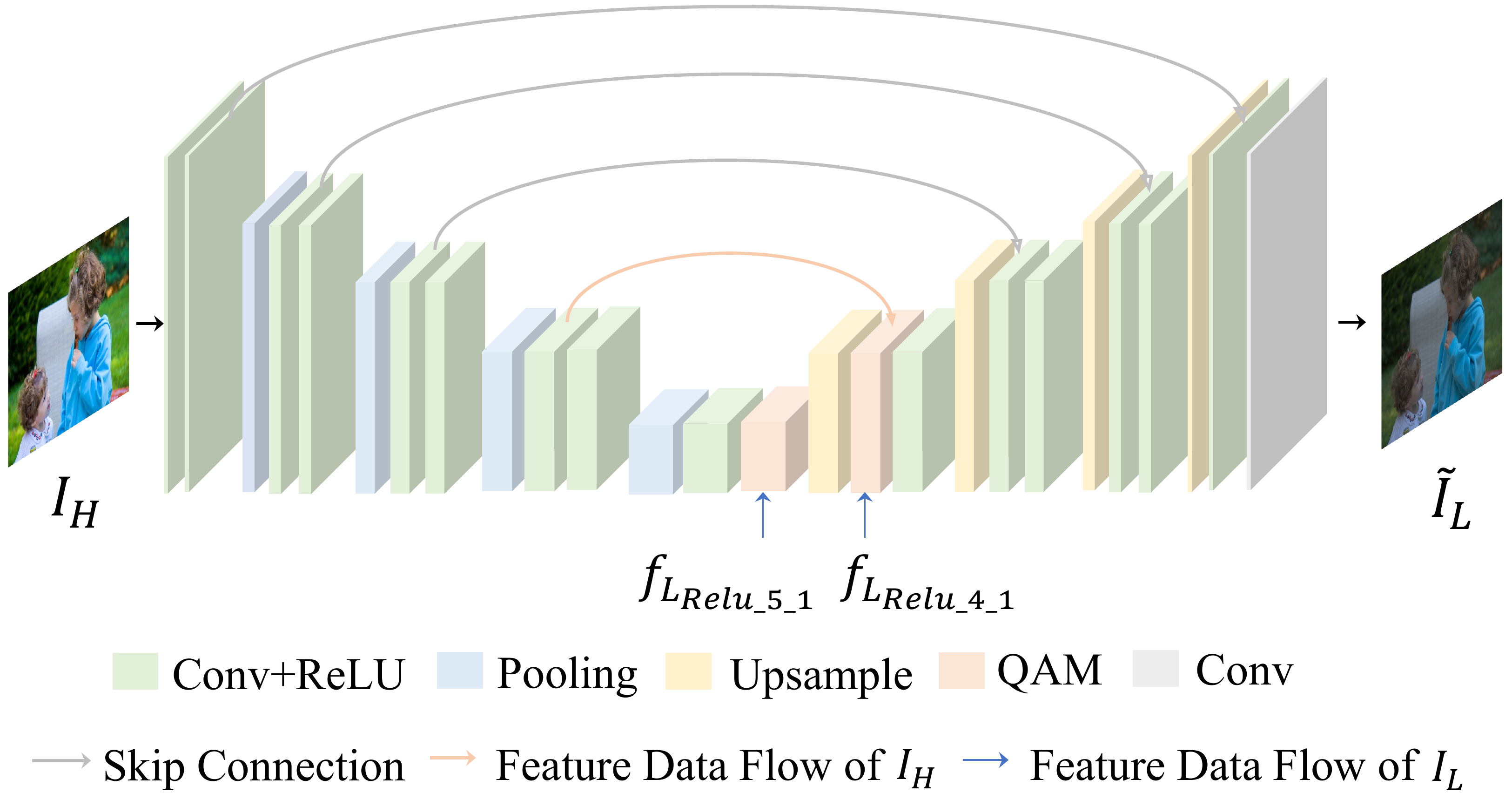}
  \caption{The structure of the encoder $E$ and decoder $G_L$. We employ the pre-trained VGG19~\cite{simonyan2014very} as the features extractor and use the features at five layers (\textit{i.e.}, Relu\_X\_1, X=1, 2, 3, 4, 5). The $f_{L_{Relu\_5\_1}}$ and $f_{L_{Relu\_4\_1}}$ are the features of $I_L$ at corresponding layers, respectively.}
\label{fig:gl}
\end{figure}

As shown in Fig.~\ref{fig:gl}, based on the pre-trained VGG~\cite{simonyan2014very}, a symmetric decoder $G_L$ and the proposed QAM, an encoder-decoder network is constructed to perform image quality degradation. We employ the pre-trained VGG~\cite{simonyan2014very} as the feature extractor and use features in five layers (\textit{i.e.}, Relu\_X\_1, X=1, 2, 3, 4, 5). The deeper the network is, the larger the receptive field size is. Therefore, we use two layers (\textit{i.e.}, Relu\_5\_1 and Relu\_4\_1) to capture the characteristics of low-quality features at different levels. To be specific, fed
\{$f_{H_{Relu\_5\_1}}$, $f_{L_{Relu\_5\_1}}$\} and \{$f_{H_{Relu\_4\_1}}$, $f_{L_{Relu\_4\_1}}$\} into the first and second QAM, respectively, to generate the quality-aware features at the highest and second-highest level, respectively.
Then fuse the two levels of quality-aware features after upsampling the spatial size of the former, which is then fed into the following modules to generate the low-quality images. The generator $G_H$ is a symmetrical decoder of VGG-19 to form U-Net~\cite{ronneberger2015u}, which is used to translate the low-quality images into the high-quality versions.

\begin{figure*}[t]
\centering
\subfloat[][Input]{\includegraphics[width=0.245\textwidth]{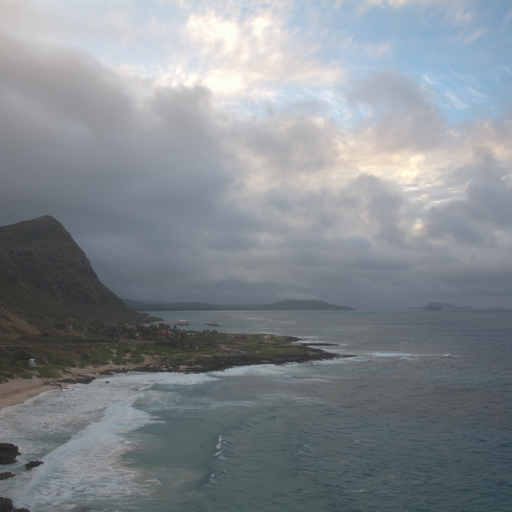}}\hfil%
\subfloat[][CycleGAN~\cite{zhu2017unpaired}]{\includegraphics[width=0.245\textwidth]{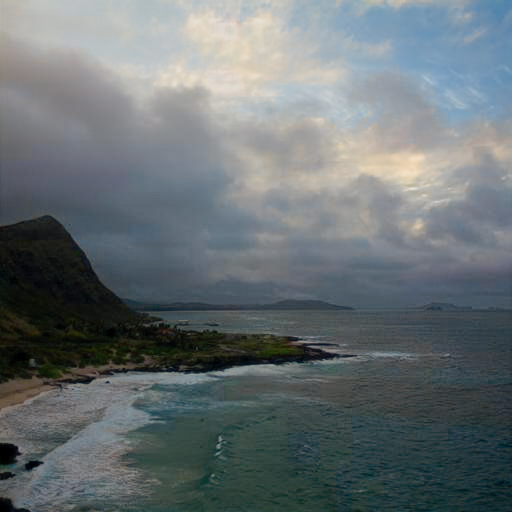}}\hfil%
\subfloat[][Exposure~\cite{hu2018exposure}]{\includegraphics[width=0.245\textwidth]{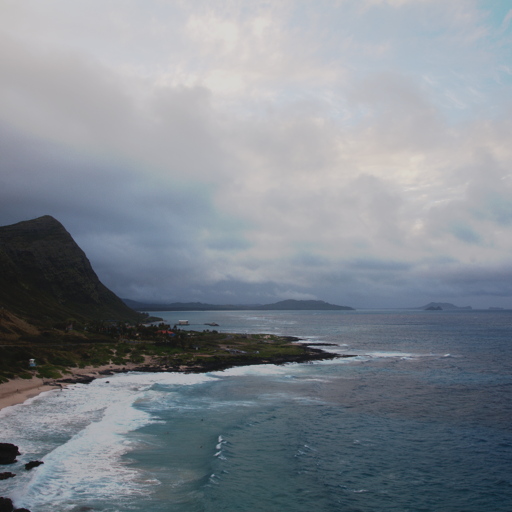}}\hfil%
\subfloat[][EnlightenGAN~\cite{jiang2019enlightengan}]{\includegraphics[width=0.245\textwidth]{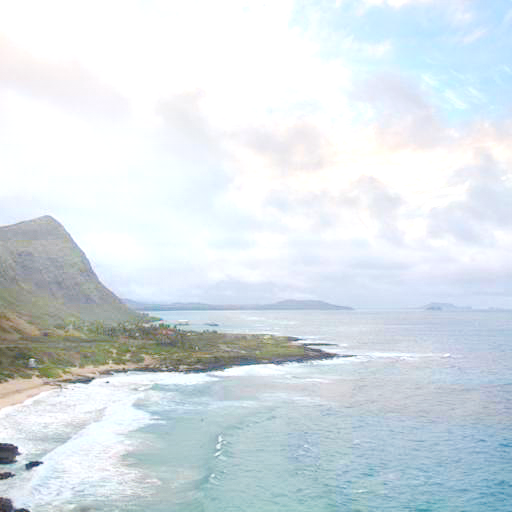}}\\
\vspace{-0.8em}
\subfloat[][DPE~\cite{chen2018deep}]{\includegraphics[width=0.245\textwidth]{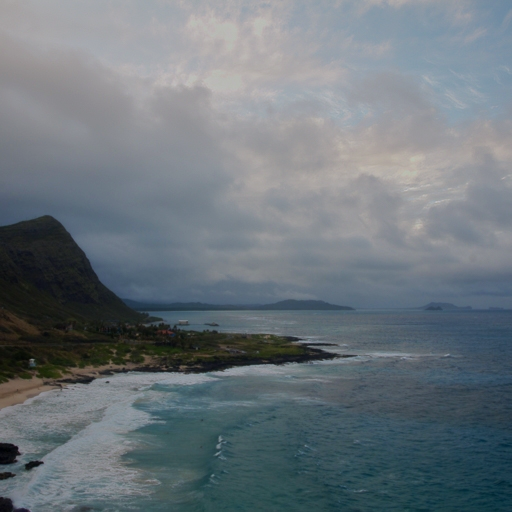}}\hfil%
\subfloat[][Ours (FiveK)]{\includegraphics[width=0.245\textwidth]{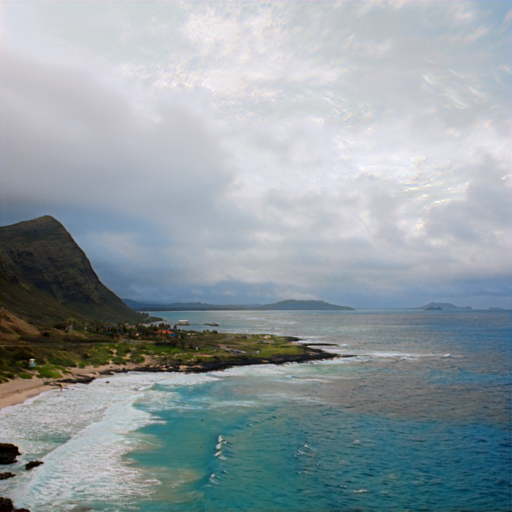}}\hfil%
\subfloat[][Ours (Flickr)]{\includegraphics[width=0.245\textwidth]{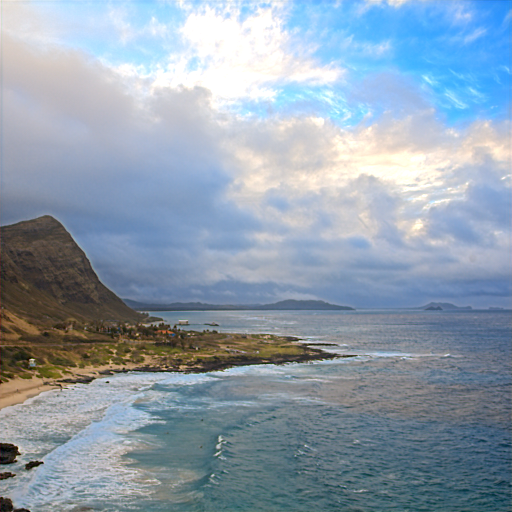}}\hfil%
\subfloat[][Expert-retouched]{\includegraphics[width=0.245\textwidth]{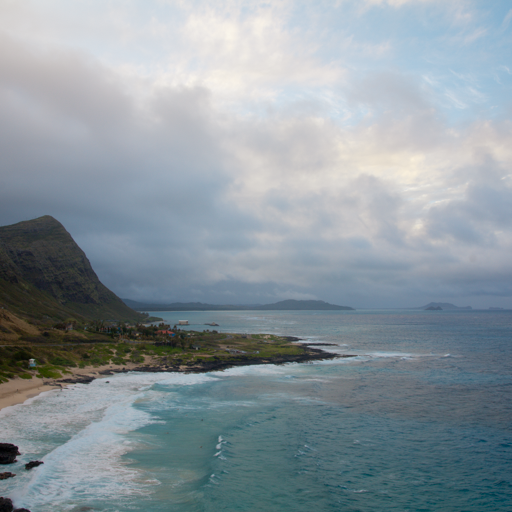}}\\
\caption{The visual quality comparison results between the proposed method against state-of-the-art methods.}
\label{fig:comparison1}
\end{figure*}

Generally speaking, a larger receptive field is beneficial for the discriminator to capture the global features to judge the real image and the generated image. Although the low-level features of the discriminator only have a smaller receptive field, it can guide the generator to produce better details. We propose a multi-scale discriminator that uses not only the final layer of the discriminator but also multiple low-level features (\textit{i.e.}, an intermediate layer of the discriminator) to guide the generator to generate the image with both finer details and global consistency.

\subsection{Loss Function}
\label{ssec:loss}

\emph{\quad 1) Adversarial Loss:} The adversarial loss used to train the mapping functions is a variation of the recently proposed relativistic discriminator structure~\cite{jolicoeur2018relativistic}. On the one hand, this loss evaluates the probability that real data is more real than fake data, on the other hand, it also enforces the generated data is more realistic than the real data. The Relativistic average HingeGAN (RaHingeGAN) of mapping from low-quality to high-quality can be expressed as follows:
\begin{equation}
\small
\begin{split}
\label{equ:hadvloss}
L^H_{\text{adv}} =
\mathbb{E}_{\tilde{I}_H\sim P_{\tilde{H}}}[\text{max}(0, 1 + (D(\tilde{I}_H) - \mathbb{E}_{I_H\sim P_{H}}D_H(I_H)))] \\
+ \mathbb{E}_{I_H\sim P_{H}}[\text{max}(0, 1 - (D(I_H)-\mathbb{E}_{\tilde{I}_H\sim P_{\tilde{H}}}D_H(\tilde{I}_H)))],
\end{split}
\end{equation}
where $I_H$ and $\tilde{I}_H$ denote the real data from the high-quality image dataset $P_H$, and the generated data from the generator $G_H$, respectively.

Analogously, the RaHingeGAN of the mapping function from high-quality to low-quality can be formulated as:
\begin{equation}
\small
\begin{split}
\label{equ:ladvloss}
L^L_{\text{adv}} =
\mathbb{E}_{\tilde{I}_L\sim P_{\tilde{L}}}[\text{max}(0, 1 + (D(\tilde{I}_L) - \mathbb{E}_{I_L\sim P_{L}}D_L(I_L)))] \\
+ \mathbb{E}_{I_L\sim P_{L}}[\text{max}(0, 1 - (D(I_L)-\mathbb{E}_{\tilde{I}_L\sim P_{\tilde{L}}}D_L(\tilde{I}_L)))],
\end{split}
\end{equation}
where $I_L$ and $\tilde{I}_L$ denote the real data from the low-quality image dataset $P_L$, and the generated data from the generator $G_L$, respectively.

\emph{2) Cycle-Consistency Loss:} The cycle-consistency loss is based on the assumption that an image should be mapped back to the original image after two opposite mappings. The cycle-consistency loss defined as the $\ell_2$ norm between the feature maps of the input images $I_H$ (or $I_L$) and those of the recycled images $\hat{I}_H$ (or $\hat{I}_L$) extracted by the pre-trained VGG network~\cite{simonyan2014very}, as follows:
\begin{equation}
\small
\begin{split}
\label{equ:lcc}
L_{\text{cc}} = &\frac{1}{J}\sum\limits_{j=1}^J\{ \mathbb{E}_{I_H\sim P_{H}}[\|\phi_j(I_H) - \mathbb{E}_{\hat{I}_H\sim P_{\hat{H}}}\phi_j(\hat{I}_H)\|_2] \\
&+ \mathbb{E}_{I_L\sim P_{L}}[\|\phi_j(I_L) - \mathbb{E}_{\hat{I}_L\sim P_{\hat{L}}}\phi_j(\hat{I}_L)\|_2] \},
\end{split}
\end{equation}
where $\phi_j(\cdot)$ means to extract the feature maps of the $j^{th}$ layer of the VGG network and $J$ is the total number of layers used. Specifically, we use VGG-19 layers $Relu\_1\_1$, $Relu\_2\_1$, $Relu\_3\_1$, $Relu\_4\_1$, and $Relu\_5\_1$, to compute cycle-consistency loss.

\emph{3) Total Loss:} By jointly considering the \textit{adversarial loss} and~\textit{cycle-consistency loss}, we define the final loss as the weighted sum of these losses, as follows:
\begin{equation}
\label{equ:perceptualloss}
L_{\text{total}} = L^H_{\text{adv}} + L^L_{\text{adv}} + \lambda_{\text{cc}}L_{\text{cc}},
\end{equation}
where $\lambda_{\text{cc}}$ is a weighting parameter to balance the relative importance of $L_{\text{cc}}$.

\section{Experiments}
\label{sec:results}

In this section, extensive quantitative and qualitative experimental results are presented and discussed. Specifically, we first introduced two datasets for evaluating our proposed method and the methods being compared, and described the implementation details. Then, we compared our proposed model with state-of-the-art methods in terms of subjective and objective metrics. Last, we carried out user study to consolidate the conclusions of our subjective and objective evaluations.

\begin{figure*}[htp]
\centering
\subfloat[][Input]{\includegraphics[width=0.245\textwidth]{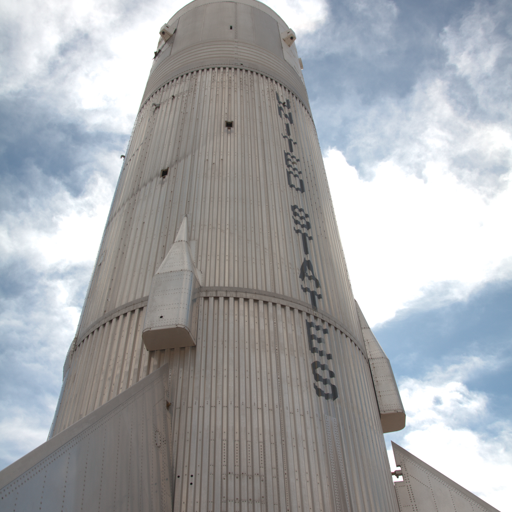}}\hfil%
\subfloat[][CycleGAN~\cite{zhu2017unpaired}]{\includegraphics[width=0.245\textwidth]{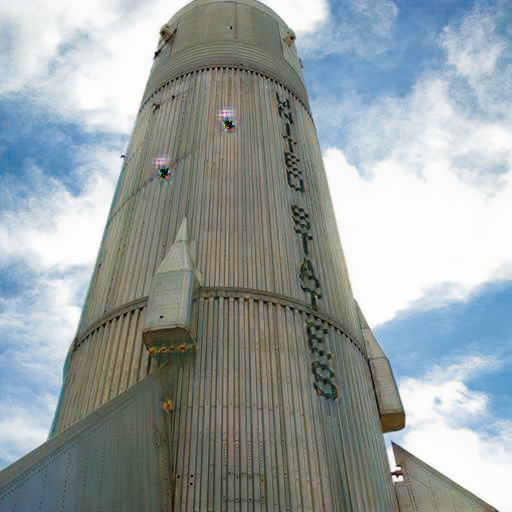}}\hfil%
\subfloat[][Exposure~\cite{hu2018exposure}]{\includegraphics[width=0.245\textwidth]{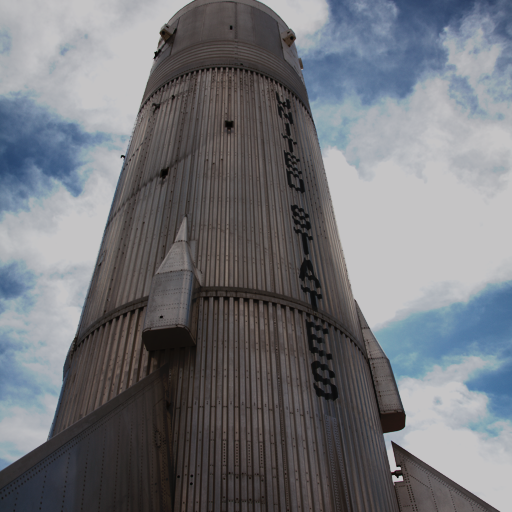}}\hfil%
\subfloat[][EnlightenGAN~\cite{jiang2019enlightengan}]{\includegraphics[width=0.245\textwidth]{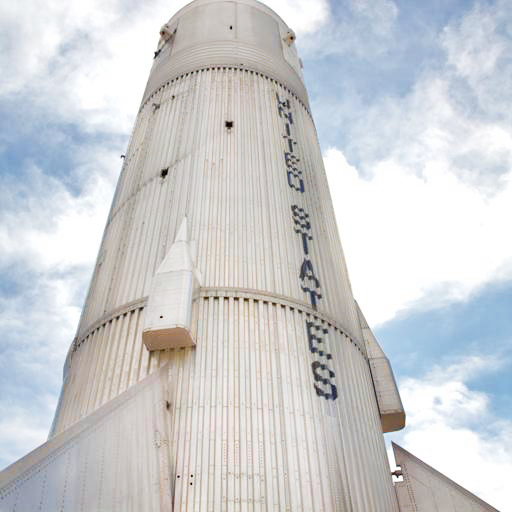}}\\
\vspace{-0.8em}
\subfloat[][DPE~\cite{chen2018deep}]{\includegraphics[width=0.245\textwidth]{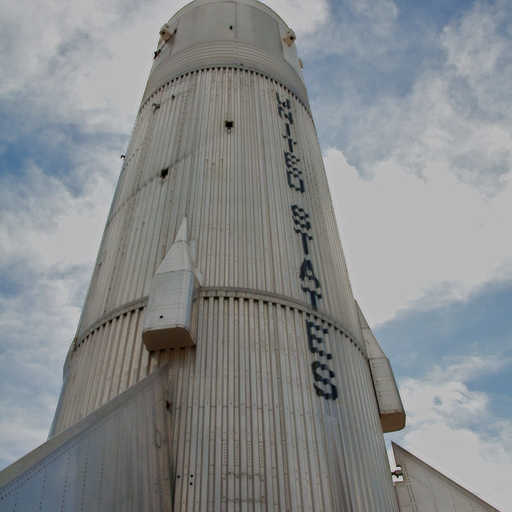}}\hfil%
\subfloat[][Ours (FiveK)]{\includegraphics[width=0.245\textwidth]{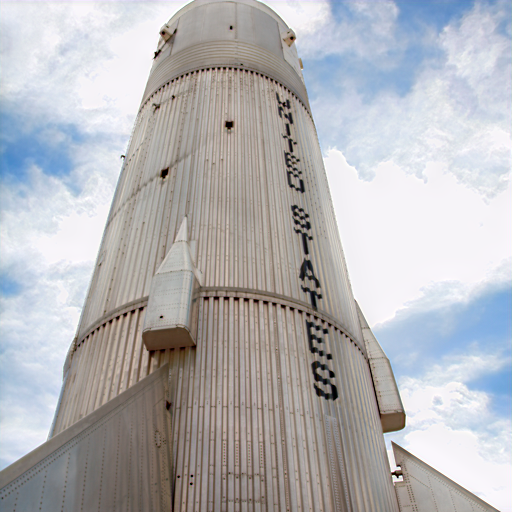}}\hfil%
\subfloat[][Ours (Flickr)]{\includegraphics[width=0.245\textwidth]{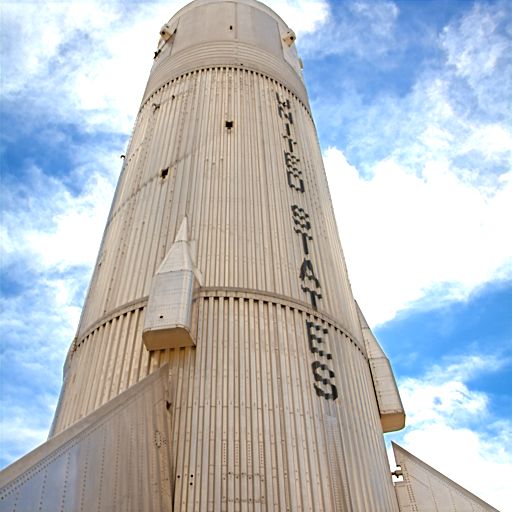}}\hfil%
\subfloat[][Expert-retouched]{\includegraphics[width=0.245\textwidth]{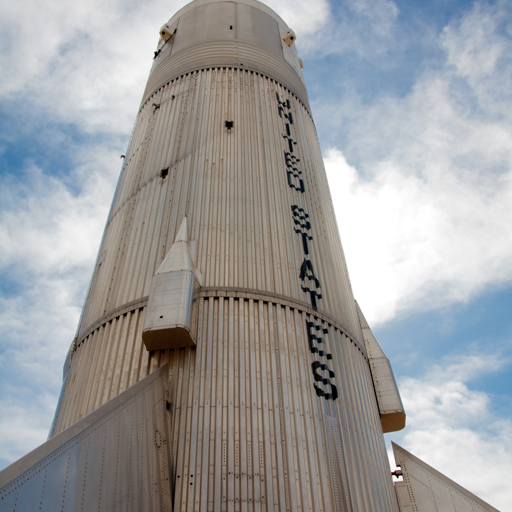}}\\
\caption{The visual quality comparison results between the proposed method against state-of-the-art methods.}
\label{fig:comparison2}
\end{figure*}

\subsection{Datasets and Evaluation Metrics}

\textit{\quad MIT-Adobe FiveK Dataset~\cite{bychkovsky2011learning}:} It consists of 5,000 raw photos taken by different photographers using various single-lens reflex cameras and 25,000 retouched photos. These retouched photos were generated from the raw photos by 5 experienced retouchers (\textit{i.e.}, retoucher A, B, C, D, and E) and the results obtained by each retoucher have different visual experiences. According to~\cite{chen2018deep} and~\cite{hu2018exposure}, the photos generated by retoucher C were selected as the target photos because his results are the most popular with users. In order to meet the requirement of the proposed model for unpaired learning, the dataset was randomly divided into three non-overlapping subsets: 1) Subset 1 is a low-quality image subset consisting of 2,250 raw photos; 2) Subset 2 is a target high-quality image subset composing of 2,250 retouched photos of other raw photos; 3) Subset 3 contains the remaining 500 raw photos used for validation (100 images) and testing (400 images).

\textit{Flickr Dataset:} In order to verify the performance of the proposed model, a high-quality image dataset crawled from Flickr to perform image enhancement under the supervision without paired data. The dataset contains 2,000 images with relatively high quality and served as the target images in unpaired training.

\textit{Evaluation Metrics:} The widely used Peak Signal to Noise Ratio (PSNR) and Structural SIMilarity (SSIM) were used to evaluate the performance of each model in terms of~\textit{signal fidelity}. In addition, the NIMA~\cite{talebi2018nima} was adopted to quantify and compare the~\textit{aesthetic quality} of the results of each model.

\subsection{Implementation Details}

We implemented our model in Pytorch and used the Adam solver~\cite{kingma2014adam} to optimize the network with $\beta_1=0.5$ and $\beta_2=0.999$. The parameters of encoder (\textit{i.e.}, VGG-19) were fixed, two generators and two discriminators were trained from scratch. The experiments were conducted on an NVIDIA GeForce RTX 2080 Ti GPU. The network was trained for 150 epochs and the mini-batch size was set as 12. The learning rates of generators and discriminators are both initialized to 1e-4 and then linearly decays to zero from 75 to 150 epochs. During training, we randomly cropped $256 \times 256$ patches and then resize them to $128 \times 128$. For testing, the resolution of all test images is $512 \times 512$. The weight parameter $\mathcal{L}_{\text{cc}}$ was empirically set to 2 and 10 for MIT-Adobe FiveK Dataset and Flickr Dataset, respectively.

\begin{table}[t]
\footnotesize
\renewcommand{\arraystretch}{1.6}
\caption{Quantitative comparison results between our proposed method and state-of-the-art methods on MIT-Adobe FiveK Dataset~\cite{bychkovsky2011learning}.}
\label{tab:quantitative}
\tabcolsep0.5cm
\centering
\begin{tabular}{l|c|c|c}
\toprule
Method  &  PSNR  &    SSIM &   NIMA \\
\midrule
Input                                 &  17.21    &  0.8012    &  4.41   \\
CycleGAN~\cite{zhu2017unpaired}       &  20.53    &  0.7786    &  4.26    \\
Exposure~\cite{hu2018exposure}       &  19.59    &  0.8361    &  4.55    \\
UIE~\cite{kosugi2019unpaired}    &  21.80       &  0.8708      &  4.53  \\
EnlightenGAN~\cite{jiang2019enlightengan} &  16.83 &  0.7513   &  4.18    \\
DPE~\cite{chen2018deep}               &  22.28    &  0.8542    &  4.39    \\
Ours  &  \textbf{22.59}    &  \textbf{0.8782}      &  \textbf{4.58}    \\
\bottomrule
\end{tabular}
\end{table}

\subsection{Quantitative Comparison}

We evaluated the performance of the proposed model against four state-of-the-art image enhancement methods learned without paired data: Deep Photo Enhancer (DPE)~\cite{chen2018deep}, EnlightenGAN~\cite{jiang2019enlightengan}, Exposure~\cite{hu2018exposure}, and UIE~\cite{kosugi2019unpaired}. We also compared it with the classical unpaired image-to-image translation model CycleGAN~\cite{zhu2017unpaired}. The DPE is based on CycleGAN model that trained with several improvements to stabilize the adversarial training process, while EnlightenGAN is a unidirectional GAN model that performs unpaired image enhancement. The Exposure and UIE are both unpaired image enhancement method based on differential filters and RL. In order to make a fair comparison, all codes of the comparison models are download from the website provided by the corresponding authors. All models under comparison are trained on the MIT Adobe FiveK dataset and our proposed model also trained on the Flickr dataset.

Table~\ref{tab:quantitative} tabulates the PSNR, SSIM, and NIMA comparison results of the proposed method against the state-of-the-art models on MIT Adobe FiveK dataset. The first-ranked performance for each evaluation metric is highlighted in bold black. For the Flickr dataset we collected, we only performed visual quality comparison and user study due to the lack of ground truth. It can be seen from Table~\ref{tab:quantitative} that on the MIT Adobe FiveK dataset, our proposed method has achieved the best PSNR, SSIM, and NIMA among all these comparison methods. Considering that DPE which ranks second in all metrics, although is the improved version of the CycleGAN from three aspects, it is still inferior to our proposed QAGAN model. We attribute this to the proposed QAM such that the generator can effectively learn domain-relevant features directly from both two domains. In addition, it can be observed that for some images the performances of CycleGAN and EnlightenGAN are even worse than the input images. This is mainly due to the CycleGAN introduces blocking artifacts, and EnlightenGAN significantly over-adjusts the contrast of images.

\begin{table}[t]
\footnotesize
\renewcommand{\arraystretch}{1.6}
\caption{The results of pairwise comparison in user study. EG denotes EnlightenGAN.}
\label{tab:userstudy}
\tabcolsep0.06cm
\centering
\begin{tabular}{lcccccccc}
\toprule
\multirow{2}{*}{}  &  \multirow{2}{*}{Input}  &    CycleGAN &   Exposure  &  EG     &   DPE   &    Ours  &  Ours  &   \multirow{2}{*}{Total} \\
    &   &  \cite{zhu2017unpaired} & \cite{chen2018deep}  & \cite{jiang2019enlightengan}  &  \cite{hu2018exposure} & (FiveK)  & (Flickr)  &   \\
\midrule
Input           & -  & 328  & 125  & 145  & 46  & 26  & 14 &  684   \\
CycleGAN           & 632  & -  & 157  & 224  & 91  & 73  &  46 & 1223    \\
Exposure          & 835  & 803   &  -  & 588  & 254  &  227  & 176  &  2883     \\
EG            & 815  & 736  & 372  & -  & 179  & 152  &  121 & 2375    \\
DPE          & 914  & 869  & 706 & 781  &  - & 362  & 267 & 3899   \\
Ours (FiveK)            & 934  & 887  & 733  & 808  & 598  & -  & 389 & 4349   \\
Ours (Flickr)     & 946  & 914  & 784  & 839  & 693  &  571  & - & 4747   \\
\bottomrule
\end{tabular}
\end{table}

\subsection{Qualitative Comparison}

We also conducted extensive visual quality comparisons between our proposed method and the state-of-the-art methods. As shown in Fig.~\ref{fig:comparison1} and Fig.~\ref{fig:comparison2}, one can observe that our proposed model trained on the collect Flickr dataset achieves the best visual experience as it produces better contrast and brightness. In addition, our proposed model trained on MIT-Adobe FiveK dataset can also produce satisfactory results. However, the methods under comparison have more or less failed in terms of contrast, color, or details. CycleGAN generates severe blocking artifacts in smooth regions and produces checkerboard artifacts at the boundary. Exposure tends to generate over-saturated results and make the colors dull. EnlightenGAN is less effective in producing good contrast and usually causes the input images to be over-exposed. DPE can produces the most comparable results to ours, but it is inferior to ours in contrast. Generally speaking, the results generated by our proposed method have comfortable contrast, vivid colors, and clear details, which are more satisfying and superior to the state-of-the-art methods in comparison.

\subsection{User Study}

It is a challenge to evaluate image quality enhancement from an objective perspective. To this end, we studied how users preference the results of each model. Specifically, we used pairwise comparison on six methods (including two versions of our method) to conduct a user study with 24 participants and 40 images. Each time, we randomly present the results of two different models to the participants and ask them to choose his/her favorite result from the displayed image pair. The results of the pairwise comparison are documented in Table~\ref{tab:userstudy}, where each value represents the number of times the method in that row outperforms the method in that column. From the results, we can observe that the results generated by the proposed method is more favourite with users, compared with other models (\textit{i.e.}, CycleGAN, Exposure, EnlightenGAN, and DPE) our proposed model has been selected the most times.
This confirms that our proposed method is not only superior to the state-of-the-art methods in the objective aspects of PSNR, SSIM, and NIMA, but is also favored by users in subjective aspect of visual quality.

\begin{figure}[t]
  \centering
  \centerline{\includegraphics[width=0.98\linewidth]{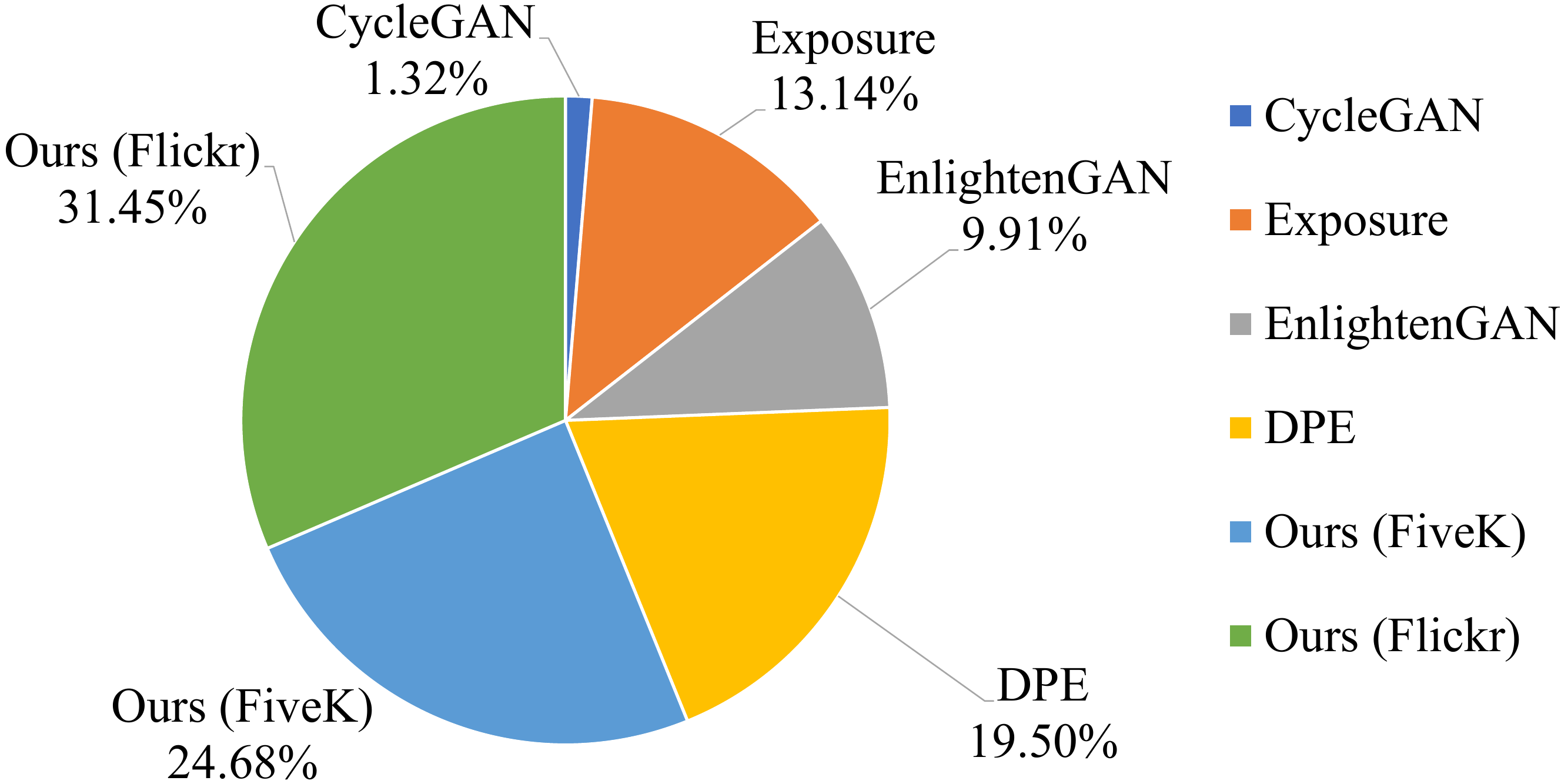}}
  \caption{User preference results of different aesthetic quality enhancement algorithms.}
\label{fig:userstudy}
\end{figure}

We also evaluated the overall quality by again randomly selecting 100 test images and 100 enhancement results corresponding to each model. Each time, six results were randomly presented to the participants and asking them to select their favorite one. At last, a total of 2400 votes were collected and visualized in Fig.~\ref{fig:userstudy}. It can be seen that the results generated by our proposed method are again preferred much more often than those by other methods under comparison.

\section{Conclusions}
\label{sec:conclusion}

In this paper, we present a bidirectional~\textit{Generative Adversarial Network} (GAN) with~\emph{quality attention module} (QAM) to train which is trained on unpaired data to solve the task of image enhancement. We embed the proposed QAM in the generator so that the generator directly learns the domain-relevant quality features from the features of low-quality images. Specifically, the proposed QAM allows the generator to effectively select semantic-relevant characteristics from the spatial-wise and adaptively incorporate style-related attributes from the channel-wise, respectively. The quantitative and qualitative experimental results show that our proposed method is superior to the state-of-the-art methods.

\begin{acks}
The authors would like to thank the anonymous referees for their insightful comments and suggestions. This work was supported in part by the Hong Kong RGC General Research Funds under Grant 9042322 (CityU 11200116), Grant 9042489 (CityU 11206317), and Grant 9042816 (CityU 11209819), and in part by the Natural Science Foundation of China under Grant 61672443.
\end{acks}


\bibliographystyle{ACM-Reference-Format}
\balance
\bibliography{QAGAN}

%
%
%
%
%
%
%
%
%
%

\end{document}